\documentstyle[11pt]{article}
\newcommand{\be}{\begin{equation}}
\newcommand{\ee}{\end{equation}}
\begin{document}
\title{Selftrapping and Quantum Fluctuations in the Discrete Nonlinear
Schr\"{o}dinger Equation\footnote{Invited talk to the {\em International Workshop on
Disordered Systems with Correlated Disorder}, Universidad de Tarapac\'{a},
Arica, August 31-September 5, 1998}}
\author{{\bf C.A. Bustamante and M. I.  Molina}
\vspace{1 cm}
\and
\and
Facultad de Ciencias, Departamento de F\'{\i}sica, Universidad de Chile\\
Casilla 653, Las Palmeras 3425, Santiago, Chile.\\
mmolina@abello.dic.uchile.cl}
\date{}
\maketitle
\begin{center}
{\bf Abstract}
\end{center}
\noindent
We study the robustness of the selftrapping phenomenon exhibited by
the Discrete Nonlinear Schr\"{o}dinger (DNLS) equation against
the effects of nonadiabaticity and quantum fluctuations in a
two--site system (dimer). To test for nonadiabatic
effects (in a semiclassical context), we
consider the dynamics of an electron (or excitation) in a dimer system 
and coupled to the vibrational degrees of freedom, modeled here
as classical Einstein oscillators of mass $M$. For relaxed
(coherent state) oscillators initial condition, the DNLS selftrapping
transition persists for a wide range of $M$ spanning 5 decades.
When undisplaced initial conditions are used, the selftrapping
transition is destroyed for masses greater than $M \sim 0.02$.
To test for the effects of
quantum fluctuations, we performed a first-principles numerical 
calculation for the fully quantum version of the above system:\
the two--site Holstein model. We computed the long-time averaged
probability for finding the  electron
at the initial site as a function of the asymmetry and nonlinearity
parameters, defined in terms of the electron-phonon coupling strength
and the
oscillator frequency. Substantial departures from the usual DNLS
system are found:\ A complex landscape in asymmetry--nonlinearity
phase space, which is crisscrossed with narrow ``channels'',
where the average electronic probability on the initial site remains
very close to $1/2$, being substantially larger outside.
In the adiabatic case, there are also low--probability ``pockets'' where
the average electronic probability is substantially smaller than
$1/2$.

\vspace{1cm}

\noindent
PACS:\ \ 71.38.+i,\ 63.20.kr,\ 63.20.pw

\newpage
\section{Introduction}

\noindent
The Discrete Nonlinear Schr\"{o}dinger (DNLS) equation is a paradigmatic
equation that describes properties of chemical, optical or condensed
matter systems where selftrapping mechanisms are present. It can
describe the dynamics of a set of anharmonic oscillators\cite{scott},
the power exchange between  nonlinear optical couplers\cite{deering}
or the motion of a quantum mechanical particle propagating in a discrete
medium while interacting strongly with vibrations\cite{mt_prl}. In this
last case, for an electron (or excitation) propagating in a discrete one-dimensional medium,
DNLS has the form
\be
i {dC_{n}\over{dt}} = V (C_{n+1} + C_{n-1}) - \chi \, |C_{n}|^{2} \, C_{n}
\label{eq:1}
\ee
where $C_{n}(t)$ is the probability amplitude of finding the electron
on site $n$ at time $t$, $V$ is the nearest-neighbors
hopping term and $\chi$ is the nonlinearity parameter, proportional
to the square of the (strong) electron-phonon coupling on site $n$.
The DNLS equation can be ``derived'' as the antiadiabatic limit of the
problem of a quantum excitation propagating in a lattice and coupled to the
vibrational degrees of freedom of each site, modelled as a set of
classical Einstein oscillators
\be
i {dC_{n}\over{dt}} =  V (C_{n+1} + C_{n-1}) + \alpha \, u_{n} \, C_{n}\label{eq:2}
\ee
\be
M\, \ddot{u}_{n} + k\, u_{n} = -\alpha\, |C_{n}|^{2}\label{eq:3}
\ee
After the transients, the oscillator amplitude $u_{n}(t)$ is given by
$u_{n} = (1/\sqrt{M\, k}) \int_{0}^{t} dt' \, \sin(\omega (t - t') \, 
\alpha \, |C_{n}(t')|^{2}$.
If we assume the oscillator time scale much shorter than the electronic
time scale for hopping, then we have approximately,
$u_{n} \approx (\alpha/k) \, |C_{n}(t)|^{2}$. Inserting this
into (\ref{eq:2}), leads to the DNLS equation (\ref{eq:1}), with $\chi =
\alpha^{2}/k$.
Perhaps the most striking feature of DNLS is that it leads to
{\em selftrapping}:\ for nonlinearity parameter values exceeding certain
threshold (whose exact value depends on the details of the initial
electron state), a portion of the electron remains bound to a particular site(s),
while the rest propagates away  in a ballistic manner \cite{mt_prl}. The
effect is most noticeable when completely localized initial conditions are
used.

The ability to partially trap an electron, suggests a possible mechanism for
tuning the electrical conductivity. It is then clear that the robustness
of the selftrapping phenomenon exhibited by the DNLS equation, against
perturbing  effects commonly found in realistic situations must be
ascertained in order to implement possible applications. In this work we
focus on two effects:\ finite inertia and quantum fluctuations. In order
to keep the discussion simple, we will work on the smallest nontrivial
DNLS unit:\ the nonlinear dimer\cite{optics}. 

This paper is organized as follows: In section 2 we review the DNLS equation
for the symmetric nonlinear dimer and its selftrapping phenomenology. We also
consider a generalized nonlinear dimer. In section 3, we introduce finite
oscillator inertia and examine its consequences on selftrapping. In section 4
we examine the fully quantum mechanical problem, given by the two-site
Holstein model. We further specialize to the symmetric (section 4.1) and
general (section 4.2) cases focussing on selftrapping properties. We end
with a brief discussion of the preliminary results found so far.

\section{DNLS dimer}
In the case of two sites, system (\ref{eq:1}) reduces to
\begin{eqnarray}
i {dC_{1}\over{dt}} & = & V \, C_{2} - \chi \, |C_{1}|^{2} \, C_{1} \nonumber\\
i {dC_{2}\over{dt}} & = & V \, C_{1} - \chi \, |C_{2}|^{2} \, C_{2}\label{eq:4}
\end{eqnarray}
For completely localized initial conditions $|\, C_{1}(0)\, |^{2} = 1$,
an exact time-dependent solution
for $C_{1}(t)$ and $C_{2}(t)$ can be found\cite{kc}. The solution
exhibits a {\em selftrapping transition}:\ the probability at the
initial site $|C_{1}(t)|^2$, oscillates in a periodic fashion for
$\chi < 4 V$, with a period that increases with nonlinearity. At
the critical value $\chi = 4 V$ the period becomes infinite and
for $\chi > 4 V$ the particles selftraps. The long time-averaged
probability $<\!|C_{1}|^{2}\!>$ obeys\cite{kt_prb}
\be
<|C_{1}|^{2}> = \left\{ \begin{array}{ll}
  			1/2 & \mbox{$\chi < 4 V$}\\
  			1/2\left (1+{{\textstyle \pi\over{2 K(4 V/\chi)}}}\right) & \mbox{$\chi > 4 V$}
  			\end{array}
  			\right.\label{eq:5}
\ee
where $K$ is the complete elliptic integral of the first kind. Another
closely related system, the {\em nondegenerate} nonlinear dimer,
considers an additional site energy difference in (\ref{eq:4}) and can
be similarly solved\cite{nnd} in terms of jacobian elliptic functions.
It is found that there is a transition
line in parameter space that separates the localized from nonlocalized
states. The difference with the previous (degenerate) case is that
now the selftrapped state can be accessed without crossing the transition
line.

We can pose the most general nonlinear dimer by incorporating, besides
a site energy mismatch, different nonlinearity parameters:
\begin{eqnarray}
i {dC_{1}\over{dt}} & = & \delta \, C_{1} + V \, C_{2} - \chi_{1} \,|C_{1}|^{2} \, C_{1} \nonumber\\
i {dC_{2}\over{dt}} & = & V \, C_{1} - \chi_{2} \, |C_{2}|^{2} \, C_{2}\label{eq:6}
\end{eqnarray}
The transformation
$\overline{C}_{i} = C_{i} \, \exp[\, (1/2)(\chi_{1}-\chi_{2}) \, |C_{2}|^{2} \, ]$ along
with the normalization condition $|C_{1}|^{2} + |C_{2}|^{2} = 1$ maps
(\ref{eq:4}) into
\begin{eqnarray}
i {d\overline{C}_{1}\over{dt}} & = & \Delta \, \overline{C}_{1} + V \, \overline{C}_{2} - \chi \, |\overline{C}_{1}|^{2} \, \overline{C}_{1} \nonumber\\
i {d\overline{C}_{2}\over{dt}} & = & V \, \overline{C}_{1} - \chi \, |\overline{C}_{2}|^{2} \, \overline{C}_{2}\label{eq:7}
\end{eqnarray}
where $\chi = (1/2)(\chi_{1} + \chi_{2})$ and $\Delta =
\delta - (1/2)(\chi_{1} - \chi_{2})$.
Therefore, if we are only interested in site occupation probabilities, the
general nonlinear dimer is equivalent to the one studied earlier by
Tsironis\cite{nnd}. In Fig. 1 we show a probability density phase
diagram for $<\!P\!>_{T}$ in $\Delta$--$\chi$ space. The transition line,
which starts at $\chi = 3.09$, \ $\Delta = -0.383$, is clearly visible.

\section{Semiclassical dimer:\ finite inertia effects}
As we have mentioned, the DNLS equation is obtained in the limit of a
negligible oscillator inertia in (\ref{eq:3}), where the oscillator
adjust instantaneously to the electronic presence. We now examine the
robustness of the selftrapping phenomenon when finite oscillator inertia
is taken into account.

While in the limits of both small and large mass,
analytical work is possible via perturbation theory\cite{chen}, we
numerically integrate the dynamical equations for the full dimer system,
equations (\ref{eq:2}) and (\ref{eq:3}) with $n = 1, 2$.
The electron is placed initially 
completely on site $1$:\ $C_{1}(0) = 1$. We explore two possible
oscillator initial conditions:\ {\em relaxed} type, where oscillator $1$
starts from rest with a displacement determined assuming a complete
oscillator relaxation $u_{1}(0) = (\alpha/ k) \,
|C_{1}(0)|^{2} = \alpha/k, \dot{u}_{1}(0) = 0$, and {\em undisplaced}, where
$u_{1}(0) = 0 = \dot{u}_{1}(0)$. For oscillator 2, $u_{2}(0) = 0 =
\dot{u}_{2}(0)$ in both cases. The coupled equations are solved using a
fourth-order Runge-Kutta scheme, whose precision is monitored through
the conservation of probability (norm).

We examine the effect of nonlinearity on the
time-averaged probability at the initially occupied site
$ < P >_{T} = (1/T) \int_{0}^{T} |C_{1}(t)|^{2} dt$
(with $T\gg 1/V,\ \sqrt{M/k}$), as an indicator for the occurrence of
selftrapping. Results are shown in Fig. 2. We note that, for relaxed
initial conditions, the finite inertia of the oscillators plays
practically no role in the existence and location of the selftrapping
transition, for at least five decades in $M$. There is a slight shift
in the critical nonlinearity towards smaller values, as $M$ is
increased. On the contrary, for
undisplaced initial conditions, the selftrapping transition is obliterated
when $M$ exceeds $0.02$ approximately. Since relaxed initial conditions are
the ``natural'' ones when the oscillators have small mass, we conclude that
the DNLS selftrapping transition is robust against finite
inertia effects. Similar conclusions have been reached in the case of 
a number of finite vibrational impurities embedded in a linear
chain\cite{chen}.

\section{Quantum dimer}
We come now to the most delicate effect:\ quantum fluctuations. The
DNLS equation was ``derived'' for the coupled electron-phonon
system by making two approximations. The oscillators were first treated
classically. Later, an antiadiabatic limit was taken to effectively
eliminate the vibrational degrees of freedom, which become completely
correlated (enslaved) with the electronic amplitude.

In this section,
we drop the previous assumptions and perform a fully quantum mechanical
calculation for the rather general two-site Holstein model\cite{holstein}
\be
H  =  H_{\mbox{\small el}} + H_{\mbox{\small ph}} + H_{\mbox{\small e-ph}} \label{holstein}
\ee
with
\begin{eqnarray}
H_{\mbox{\small el}}  & = & \delta \, C_{1}^{\dagger} C_{1} - V (C_{1}^{\dagger} C_{2} +
	  C_{2}^{\dagger} C_{1})\\
H_{\mbox{\small ph}}   & = & \omega\, (a_{1}^{\dagger} a_{1} + a_{2}^{\dagger} a_{2})\\
H_{\mbox{\small e-ph}} & = &  g_{1} \, C_{1}^{\dagger} C_{1} (a_{1}^{\dagger} + a_{1}) +
           g_{2} \, C_{2}^{\dagger} C_{2} (a_{2}^{\dagger} + a_{2})
\end{eqnarray}
where $\delta$ is a site energy asymmetry, $C_{j}^{\dagger}(C_{j})$ creates (destroys) an electron on the
$j$th site, $a_{j}^{\dagger} (a_{j})$ is the usual phonon creation
(destruction) operator at site $j$, $\omega$ is the (Einstein) oscillator
frequency, chosen to be the same for both oscillators for simplicity and
$g_{j}$ is the electron-phonon coupling at site $j$. The nonlinearity
parameter used in DNLS corresponds here to $\chi_{j} = g_{j}^{2}/\omega$.
Previous work on a single Holstein impurity embedded in
a finite chain have displayed significant departures from the DNLS
behavior\cite{pla}. In that case, it was proved that in the antiadiabatic
limit, the Hamiltonian becomes equivalent to the one of a static, linear
impurity and $<\!P\!>_{T}$ never converges to the DNLS profile.

We decouple one of the phonon fields in (\ref{holstein}) using the canonical transformation
\begin{eqnarray}
a & = & {1\over{G}} \, ( \, g_{1}\ a_{1} - g_{2}\ a_{2} - {g_{2}^{2}\over{\omega}}\, )\nonumber\\
b & = & {1\over{G}} \, ( \, g_{2}\ a_{1} + g_{1}\ a_{2} + {g_{1}\ g_{2}\over{\omega}})
\end{eqnarray}
where $G \equiv (\, g_{1}^{2} + g_{2}^{2}\, )^{1/2}$ is an effective coupling.
Using $\Delta \equiv \delta + (g_{2}^{2} - g_{1}^{2})/\omega$ as the
asymmetry parameter, the transformed Hamiltonian reads
\be
H = (\Delta + G^2/\omega)  C_{1}^{\dagger} C_{1} + V\, (C_{1}^{\dagger} C_{2} +
C_{2}^{\dagger} C_{1} ) + \omega \, a^{\dagger} a + G\, C_{1}^{\dagger} C_{1} ( a + a^{\dagger})
+ (\omega\, b^{\dagger} b - g_{2}^{2}/\omega) 
\label{eq:H}
\ee
The Hamiltonian thus simplified allows for an exact numerical
diagonalization scheme, where the eigenenergies and quantum
amplitudes are computed by expanding in a phonon
basis set\cite{carlos}. With the eigenenergies and eigenfunctions we
are ready to compute dynamical observables. As the initial condition
for the electron we choose the one that places it completely on one
of the sites (``site 1'') and focus on the probability $P(t)$ for
finding it there at an arbitrary time later. As before, 
we shall pay particular attention to the long-time average of $P(t)$,
$<\!P\!>_{T} = (1/T)\, \int_{0}^{T} P(t)\, dt$, as well as
$<\!P\!>_{\infty} = \lim_{T\rightarrow \infty} <\!P\!>_{T}$. For
the phonon part, we use two different initial conditions:
undisplaced oscillators (i.e., zero phonons present) and oscillators 
naturally ``relaxed'' with the electron presence (a coherent state).

\subsection{Symmetric quantum dimer}

Let us consider the special situation $\Delta = 0$. In this case it is possible
to prove that $<\!P\!>_{\infty} = 1/2$, independent on the initial
conditions\cite{carlos}. The original Hamiltonian can be separated into
two diagonal blocks $H_{\pm}$ with eigenvalues
$\{\epsilon_{\pm}\}$ and eigenvectors $\{|\epsilon_{\pm}\!>\}$. The
probability $P(t)$ reads
\be
P(t) = (1/2) + \sum_{\epsilon_{+},\epsilon_{-}} C(|\epsilon_{\pm}\!>)
\cos( (\epsilon_{+}-\epsilon_{-}) t )
\ee
where the coefficients $C(|\epsilon_{\pm}\!>)$ depend on the initial
conditions and on the eigenvectors $|\epsilon_{\pm}\!>$. Barring any
degeneracy, $<\!P\!>_{\infty} = 1/2$, regardless of the initial conditions.

In the rest of this section we will use the coupling parameter
$g \equiv G/\sqrt{2}$ since it allows a better comparison with
previous works. It corresponds to the coupling per site when $\delta =0$.

We are particularly interested in the adiabatic ($\omega \ll V$) and
antiadiabatic ($\omega \gg V$) limits. For the antiadiabatic case, we obtain
\be
P(t) \approx (1/2) (\, 1 + \cos(\ 2\ V_{\mbox{\small eff}}\ t\ ) \, )\label{veff}
\ee
where the effective hopping term $V_{\mbox{\small eff}}$ decreases with $g/\omega$
as $V_{\mbox{\small eff}} = V \, \exp(-\alpha (g/\omega)^{2})$ (with $\alpha = 0.434$
for $\omega \geq 10$). This is the well-known ``polaronic narrowing''
of the band, a limit well understood which can be obtained by first
performing a Lang-Firsov transformation on the Holstein
Hamiltonian (\ref{holstein}), followed by the $\omega/V \gg 1$ limit.
For the adiabatic regime, the behavior of the electron is similar to the
previous case only in the limits $g/\omega < 1$ or $g/\omega \gg 1$. In the
intermediate region, $P(t)$ displays a complex behavior due to the
simultaneous presence of many different frequency scales\cite{kenkre}

Even though $<\!P\!>_{\infty} = 1/2$ would seem to preclude any selftrapping
phenomenon like the one observed in DNLS, it is clear from 
(\ref{veff}), that the effective hopping
can decrease to the point of ``trapping'' the electron for
long times on the initial site. For a given observational time
scale $T$, we examine $<\!P\!>_{T}$. Results are shown in
Figs. 3a and 3b. We note that
the behavior of $<\!P\!>_{T}$ is analogous to its DNLS counterpart\cite{T},
with a ``critical'' value of $g/\omega$ above which the electron tends
to confine itself at the initial site, forming a ``small polaron'',
a localized electron with low mobility.

The transition region between free electron and a small polaron
behavior, can be determined  
by the following criterion\cite{capone}: examination of the correlation
$\chi_{1,1} = <\!\phi_{0}|\, C_{1}^{\dagger} C_{1} (a_{2}^{\dagger} +
a_{2} )\, |\phi_{0}\!>$ where $|\phi_{0}\!>$ is the ground state.
As $g/\omega$ is increased, $\chi_{1,1}$ decreases initially from zero, 
reaches a minimum value and rises again back to zero.
The maximum value of its slope $d\chi_{1,1}/d (g/\omega)$
determines the critical point $(g/\omega)_{c}$ and the maximum (minimum)
value of its second derivative $d^{2} \chi_{1,1}/  d(g/\omega)^{2}$
mark the lower (upper) boundary of the transition region. This is shown
in Fig. 3c. Other criteria, such as variations of the
ground
state energy or the examination of
$\chi_{1,0} = <\!\phi_{0}\, |\, C_{1}^{\dagger} C_{1} (a_{1}^{\dagger} + a_{1} )\, |\phi_{0}\!>$
give the same qualitative results. The features displayed in Fig. 3c
seem to remain valid also for longer chains, as shown by Capone et al.\cite{capone}
for $N = 4$ sites using exact diagonalization and by Romero et al.\cite{romero} for $N = 32$ sites, who
relies on a variational scheme to obtain $|\phi_{0}\!>$ plus an energetic
criteria. The latter group propose for the ``critical'' line the relation $(g/\omega)_{c} =
1 + (V/\omega)^{1/2}$. The agreement for the dimer case is good, as can
be seen in Fig. 3c. According to Capone et al.\cite{capone}, a criterion
for the formation of a small polaron is the simultaneous occurrence
of
\be
\left({g\over{\omega}}\right) > 1 \hspace{1cm}\mbox{and}\hspace{1cm}
		\left({g^{2}/\omega\over{2 V}}\right) > {1\over{2}}
\ee
that is, there must be a net phononic displacement $g/\omega$ correlated
with an energetic gain $g^{2}/\omega$ from polaron formation greater
than the kinetic energy ($\sim 2 V$) lost due to electron localization.

\subsection{General quantum dimer}

The general quantum dimer is described by two parameters: the asymmetry
$\Delta = \delta + (g_{2}^{2} - g_{1}^{2})/\omega$ and the effective
coupling $G = (\ g_{1}^{2} + g_{2}^{2}\ )^{1/2}$.
We focus on the values for $<\!P\!>_{\infty}$ in terms of $\Delta$ and $G$, and for the two phonon initial
conditions: undisplaced and relaxed. Results are displayed in Fig. 4 in
the form of probability density phase diagrams.

A comparative study of $<\!P\!>_{\infty}$ in terms of
$\Delta$ requires to consider two scales. One of them is the hopping
$V$. For small $G/\omega$ values, variations in $<\!P\!>_{\infty}$ are
determined by this scale. The other relevant scale is the frequency
$\omega$. When the phonons are strongly coupled to the electron
(i.e., when $G/\omega$ is large enough), the effective electronic
hopping increases when the asymmetry value is a multiple of the frequency:
$\Delta\approx n \omega$
with $n = 0,1,2,...$ for relaxed initial conditions and
$n = 0,\pm 1,\pm 2,...$
for undisplaced initial conditions. On these ``resonance'' channels,
$<\!P\!>_{\infty} = 1/2$. As $\omega$ is decreased, these channels become
narrower and begin to distort considerably the low asymmetry region
$(|\Delta| \leq 5$) giving rise to a probability pattern of
great complexity. One of its most salient
features is the presence of basins or low-probability ``pockets'' where $<\!P\!>_{\infty}$
is significantly smaller than $1/2$. These pockets appear for $\Delta >0$
and can be clearly appreciated in Fig. 5, which shows a blow-up of the density
phase diagram for the adiabatic case with relaxed initial conditions.
Inside one of the ``pockets'', the phonons
conspire to push the electron towards the other site, the one with 
lower site energy. On the contrary, for the
adiabatic case with undisplaced phonon initial conditions, no such pockets
have been found and, in fact, $<\!P\!>_{\infty}$ is always greater or equal to
$1/2$.

Figure 5 also shows a blow-up of the probability density phase diagram for
the DNLS case. We observe, near the DNLS critical line separating localized
from non-localized states, a low-probability pocket, whose origin
is uncertain at this stage.

The onset of the secondary resonances ($n \neq 0$) tends to occur
for ever increasing values of $G$ as $\omega$ decreases and, for fixed
$\omega$, as $n$ increases. The beginning of
fully consolidated channels can be used to define the position of a
critical coupling value $G_{c} = G_{c} (\omega, \Delta/\omega$). Above this
value, the average probability scales as $ <\!P\!>_{\infty} = \mbox{f}\, (\Delta/\omega)$
where f is a smooth periodic function ranging from $1/2$ to $1$.
We believe this is the small polaron region for the general case. A
finite-time average $<\!P\!>_{T}$ of order unity in this region
would confirm this conjecture.
This would imply a DNLS-type behavior for $<\!P\!>_{T}$.
It remains to be determined whether the $G_{c}$ line
approaches the DNLS selftrapping line (Fig. 1). If true, DNLS could be
consider to appropriately describe the selftrapping dynamics of the
quantum case, at least qualitatively, when finite-time average is considered. We are currently endeavoring to explain and model
the resonance channels observed. They have also been seen
for the case of a single vibrational impurity in a finite linear chain\cite{pla}.
\vspace{1cm}

\noindent
One of the authors (C.A.B) acknowledges partial support from
a FONDECYT doctoral grant (project 2980033).

\newpage

\centerline{{\bf Captions List}}
\vspace{2cm}

\noindent{\bf Fig. 1 :}\ \ DNLS dimer:\ probability density phase diagram
of $<\!P\!>_{T}$ in asymmetry--nonlinearity space ($T = 300\, V^{-1}$). The
transition line begins at $\chi = 3.09$, \ $\Delta = -0.383$.
\vspace{0.2cm}

\noindent {\bf Fig. 2 :}\ \ Semiclassical dimer:\ Time-averaged probability
at the initially occupied site $< P >_{T}$ (vertical) versus nonlinearity
parameter $\alpha^{2}/k$ (horizontal) for different oscillator
masses. Left column:\ Relaxed oscillator initial conditions.
Right column:\ Undisplaced oscillator initial conditions ($T = 200\ V^{-1}$ , $k=1$).
\vspace{0.6cm}

\noindent{\bf Fig. 3 :}\ \ Symmetric quantum dimer:\ The transition
region from free electron to small polaron behavior is depicted by the
shadowed area in (c). In (a) and (b) we show representative $<\!P\!>_{T}$
plots in the adiabatic and antiadiabatic region, respectively ($T = 500\ V^{-1}$ , $V\equiv 1$).
 The thick (thin) line corresponds to initially relaxed (undisplaced) oscillators.
\vspace{0.6cm}

\noindent{\bf Fig. 4 :}\ \ General quantum dimer:\
Probability density phase diagrams for $<\!P\!>_{\infty}$ in
asymmetry--``nonlinearity'' space. Upper row:\ Adiabatic case
($\omega = 0.5$). Lower row:\ Antiadiabatic case ($\omega = 10$).
The left (right) column refers to relaxed (undisplaced) initial oscillators 
condition. ($V\equiv 1$).
\vspace{0.6cm}

\noindent{\bf Fig. 5 :}\ \ Blow-up of the probability density
phase diagrams for the semiclassical nonlinear dimer (upper) and
the general adiabatic quantum dimer (lower, $\omega=0.5$). 

\newpage

\begin{thebibliography}{9}

\bibitem{scott}
J.C. Eilbeck, P.S. Lomdahl and A.C. Scott, Physica D {\bf 16}, 318 (1985).

\bibitem{deering}
See, for instance, W.D. Deering and M.I. Molina, IEEE J. Quantum Electron.
{\bf 33}, 336 (1997) and references contained therein.

\bibitem{mt_prl}
M.I. Molina and G.P. Tsironis, Phys. Rev. Lett. {\bf 73}, 464 (1994).

\bibitem{optics}
In an optical context, each site represents a nonlinear waveguide
(usually of the Kerr-type) and the relevant independent variable is not
$t$ but $z$, the distance along the fiber. In that case, the nonlinearity
parameter is proportional to the total input power and the selftrapping
occurs in the power carried by the initial waveguide, giving rise to a
nonlinear optical switch.

\bibitem{kc}
V. M. Kenkre and D.K. Campbell, Phys. Rev. B {\bf 34}, 4959 (1985).


\bibitem{kt_prb}
V.M. Kenkre and G.P. Tsironis, Phys. Rev. B {\bf 35}, 1473 (1987).

\bibitem{nnd}
G.P. Tsironis, Ph. D. Thesis, University of Rochester (1986),
G.P. Tsironis, Phys. Lett. A {\bf 173}, 381 (1993).

\bibitem{chen}
D. Chen, M.I. Molina and G.P. Tsironis, J. Phys.: Condens. Matter {\bf 5},
8689 (1993).

\bibitem{holstein}
T. Holstein, Ann. Phys. (N.Y.) {\bf 8}, 325(1959);
J. Ranninger and U. Thibblin, Phys. Rev. B{\bf 45}, 7730 (1992).

\bibitem{carlos}
C.A. Bustamante, Ph.D. Thesis, University of Chile.

\bibitem{kenkre}
M.I. Salkola, A.R. Bishop, V.M. Kenkre and S. Raghavan, Phys. Rev. B {\bf 52}, R3824 (1995).
\bibitem{T}
For $T$ as large as $10^{4} V^{-1}$ for all the examined cases.


\bibitem{capone}
M. Capone, M. Grilli and W. Stephan, Los Alamos preprint cond-mat/9606045;
M. Capone, W. Stephan and M. Grilli, Phys. Rev. B {\bf 56}, 4484 (1997).

\bibitem{romero}
A. H. Romero, David Brown and Katja Lindenberg, Los Alamos preprints cond-mat/ 9806031 -
9806032.

\bibitem{pla}
M.I. Molina, J.A. R\"{o}ssler and G.P. Tsironis, Phys. Lett. A {\bf 234},
59 (1997).

\end{thebibliography}
\end{document}